\documentclass[preprint]{elsarticle}

\usepackage[T1]{fontenc}
\usepackage[latin1]{inputenc}
\usepackage{graphicx}

\begin{document}

\begin{frontmatter}
 
\title{Improving isotopic identification with \emph{INDRA} Silicon-CsI(\emph{Tl}) telescopes}

\author[LPCC]{O.~Lopez\corref{correspondingauthor}}
\cortext[correspondingauthor]{Corresponding author}
\ead{lopezo@in2p3.fr}

\author[LPCC,HORIA]{M.~P\^arlog}
\author[IPNO]{B.~Borderie}
\author[IPNO]{M.F.~Rivet\fnref{fn}}
\fntext[fn]{Deceased}
\author[LPCC]{G.~Lehaut}
\author[IPNO,TAMU]{G.~Tabacaru}
\author[IPNO]{L.~Tassan-got}
\author[IJPAN]{P.~Paw{\l}owski}
\author[GANIL]{E.~Bonnet}
\author[LPCC]{R.~Bougault}
\author[GANIL]{A. Chbihi}
\author[NAP,IPNO]{D.~Dell'Aquila}
\author[GANIL]{J.D.~Frankland}
\author[IPNO,CNAM]{E.~Galichet}
\author[LPCC,GANIL]{D.~Gruyer}
\author[NAP]{M.~La~Commara}
\author[LPCC]{N.~Le~Neindre}
\author[LNS,NAP]{I.~Lombardo}
\author[EAMEA,LPCC]{L. Manduci}
\author[GANIL,CENBG]{P.~Marini}
\author[LPCC]{J.C. Steckmeyer}
\author[IPNO]{G.~Verde}
\author[LPCC]{E.~Vient}
\author[GANIL]{J.P. Wieleczko}
\author{ for the INDRA collaboration}

\address[LPCC]{Normandie Universit\'e, ENSICAEN, UNICAEN, CNRS/IN2P3, F-14000 Caen, France}
\address[HORIA]{Horia Hulubei National Institute for R\&D in Physics and Nuclear Engineering (IFIN-HH), P.O.BOX MG-6, RO-76900 Bucharest-M\`agurele, Romania}
\address[IPNO]{Institut de Physique Nucl\'eaire, CNRS/IN2P3, Univ. Paris-Sud, Universit\'e Paris-Saclay, F-91406 Orsay cedex, France}
\address[TAMU]{Cyclotron Institute, Texas A\&M University, MS 3366 College Station, Texas 77843, USA}
\address[IJPAN]{Institute of Nuclear Physics PAN, ul. Radzikowskiego 152, 31-342 Krakow, Poland}
\address[GANIL]{GANIL, CEA-DSM/CNRS-IN2P3,B.P. 5027, F-14076 Caen cedex, France}
\address[CNAM]{Conservatoire National des Arts et M\'etiers, F-75141 Paris Cedex 03, France}
\address[NAP]{Dipartimento di Scienze Fisiche e Sezione INFN, Universit\`a di Napoli ``Federico II'', I80126 Napoli, Italy}
\address[LNS]{INFN - Laboratori Nazionali del Sud, Via S. Sofia 62, 95125 Catania, Italy}
\address[EAMEA,LPCC]{EAMEA, CC19 F-50115 Cherbourg-Octeville Cedex, France}
\address[CENBG]{CEN Bordeaux-Gradignan, Le Haut-Vigneau, F-33175 Gradignan Cedex, France}


%

\begin{abstract}
Profiting from previous works done with the \emph{INDRA} multidetector \cite{Pouthas} on the description of the light response $\mathcal L$ of 
the CsI(\emph{Tl}) crystals to different impinging nuclei \cite{Parlog1,Parlog2}, we propose an improved $\Delta E - \mathcal L$ identification-calibration procedure for Silicon-Cesium Iodide 
(Si-CsI) telescopes, namely an Advanced Mass Estimate (\emph{AME}) method. \emph{AME} is compared to the 
usual, 
simple visual analysis of the corresponding two-dimensional map of $\Delta E - E$ type, by using \emph{INDRA} experimental data from nuclear reactions induced by 
heavy ions in the Fermi energy regime. We show that the capability of such telescopes to identify both the atomic $Z$ and the mass $A$ numbers of 
light and heavy reaction products, can be quantitatively improved thanks to the proposed approach. This conclusion opens new possibilities
to use \emph{INDRA} for studying these reactions especially with radioactive beams. Indeed, the determination of the mass for charged reaction products becomes of paramount 
importance to shed light on the role of the isospin degree of freedom in the nuclear equation of state \cite{WCI,EPJA50} .
\end{abstract}

\end{frontmatter}

\section{Introduction}

One of the present motivations for investigating heavy-ion collisions
at intermediate energies consists of improving our understanding of the
equation of state for nuclear matter with the
isospin degree of freedom. The advent of new accelerators, providing high
intensity radioactive beams will cover a broad range of isospin ($N/Z$) ratios.
Jointly, new detection arrays like FAZIA \cite{Bougault,Carboni}, which fully exploit pulse
shape analysis from silicon detectors, are under construction to
benefit from these future possibilities.
Information on the isospin dependence of the nuclear \emph{EOS} can then be obtained by properly choosing projectile-target colliding systems.
To improve the present experimental capabilities in this framework, we present a new 
Advanced Mass Estimate (\emph{AME}) approach, based upon the 
telescope technique for \emph{INDRA} Silicon-CsI telescopes
\cite{Pouthas}. This approach will extend the isotopic identification to
nuclear reaction products heavier than those commonly identified with standard $\Delta E - \mathcal L$ 
two-dimensional correlations. Here, $\Delta E$ indicates the energy lost in the $1^{st}$
Silicon stage (Si) of the telescope and $\mathcal L$ the scintillation light
produced in the $2^{nd}$ stage, made by a CsI(Tl) scintillator crystal read by a photomultiplier
and corresponds to the residual energy $E = E_0$ deposited  by 
energetic charged reaction products. The main difficulties for identifying
 the mass number over a broad range of elements are related to the non-linear energy response 
 of each of the two stages and, in particular, of the scintillator. Actually, the light response of the scintillator strongly depends on the reaction product
identity (charge and mass), which  makes difficult even the 
determination of the deposited energy. At present time, the isotopic
identification is visually achieved only for light nuclei from hydrogen up to (roughly) 
carbon isotopes for most of the \emph{INDRA} Si-CsI telescopes. 
For some specific telescopes with smaller thickness - $150~\mu m$
instead of $300~\mu m$ -, an increased gain has been used in order to 
improve the energy resolution and hence the isotopic separation during the $5^{th}$ \emph{INDRA} campaign performed at $GANIL$ a few years ago. In doing 
so, the isotopic identification for these specific telescopes has been
slightly augmented up to oxygen isotopes for the best cases.

To improve and optimize information coming from \emph{INDRA}
Si-CsI telescopes as far as the mass number is concerned, we started from the pionneering works of P\^arlog \emph{et al.} \cite{Parlog1,Parlog2} which provide an 
accurate physical description of the light response produced by the
CsI(\emph{Tl}) crystals.
In these articles, two formulas have 
been derived concerning the relation between the light signal $\mathcal L$, the atomic number $Z$, the mass number $A$ and the incident energy 
$E_0$ of a reaction product detected by a CsI
scintillator. The proposed method was then used and tested on data recorded with 
\emph{INDRA} during the fifth campaign, with telescopes having as first stage $300~\mu m$ or $150~\mu m$-thick Silicon detectors. These experimental 
data were obtained by bombarding $^{112,124}Sn$ targets with $^{124,136}Xe$ beams at $32A$ MeV and $45A$ MeV.

The paper is organized as follows. In section (II), we recall the main
results of references \cite{Parlog1,Parlog2} concerning the role of
quenching and knock-on electrons in scintillation light from the CsI(\emph{Tl}) crystals and show the quality of
the analytical description. Section (III) describes the Advanced Mass Estimate 
(\emph{AME}) method and the comparisons with standard \emph{INDRA} isotopic identification. In
section (IV) determination and uncertainty on $A$ are discussed. In Section
(V) we present a summary of this work.

\section{Quenching and knock-on electrons ($\delta$-rays) in scintillation light of CsI(\emph{Tl}) crystals}

Cesium iodide scintillators, CsI(\emph{Tl}), doped with thallium at a level of $0,02 - 0,2 \%$ molar concentration, are inorganic crystals 
 where the scintillation light is produced by the activation (excitation) of the thallium atoms encountered by the carriers (electrons and 
holes) produced during the motion of the incoming charged product. The activation  results in an emission of light by the excited thallium 
atoms in the green band at $550~nm$. The differential scintillation light output $\frac{d\mathcal L}{dE}$ as a function of energy $E$ is often described 
by means of the Birks formula \cite{Birks}:

\begin{equation}
\frac{d \mathcal{L}}{dE}=\mathcal S \frac{1}{1+\mathcal K \mathcal B\big(\frac{dE}{dx}\big)}, 
\label{equ1}
\end{equation}
$\mathcal S$ being the scintillation efficiency and $\mathcal K \mathcal B$ the quenching coefficient. The differential light decreases as 
the stopping power $\big(\frac{dE}{dx}\big)$ increases; this is the so-called quenching effect, more pronounced for the heavier ions leading to high carrier 
concentrations. Under the approximation $\big(\frac{dE}{dx}\big) \propto AZ^2/E$, the integral over the variable $E$ of the above equation 
provides a simple formula for the total light response $\mathcal{L}$ \cite{Horn} as a function of the initial energy $E_0$ of the detected ion:

\begin{equation}
\mathcal{L}(E_0) = \int_{0}^{E_{0}} \mathcal{L}(E) dE = a_1 E_0\left[1-a_2\frac{AZ^2}{E_0}\ln\big(1+\frac{1}{a_2 AZ^2/E_0}\big)\right], 
\label{equ2}
\end{equation}

The gain coefficient $a_1$ includes both the scintillation efficiency and 
the electronic chain contribution to the signal amplification. The quenching coefficient $a_2$ is mainly related to the prompt direct recombination 
of part of the electrons and holes, which thus are not participating to the excitation of the activator atoms. 

The expressions (\ref{equ1}) and (\ref{equ2}) were used, with reasonable results \cite{Birks,Horn,DeFilippo} in the case of light charged particles 
or Intermediate Mass Fragments (IMFs) of rather low energy per nucleon $E/A$, \emph{i.e.} as long as the contribution to the light response of the 
knock-on electrons or $\delta$-rays, escaping the fiducial volume of
very high carrier concentration close to the trajectory of the particle/ion, 
remains unsignificant. Actually, above a certain energy per nucleon
threshold $e_{\delta}=E_{\delta}/A$, the incident particle/ion 
starts to generate these rapid electrons, which are characterized by a small stopping power. Consequently, the fraction $\mathcal {F}(E)$ - 
firstly introduced by Meyer and Murray \cite{Meyer} -, of the energy $dE$ deposited into a slice $dx$ and carried off by the knock-on electrons 
is practically not affected by quenching. The $\delta$-rays increase thus the light output and this should be necessarily taken into account at 
energies higher than a few MeV/nucleon, especially for heavier ions.

As it penetrates into a CsI crystal, an energetic charged
particle/ion is gradually loosing its energy (from $E_0$ to $0$) mainly by ionization 
- the electronic stopping power -, leading to the scintillation, but also, in a smaller extent, by interacting with the host lattice nuclei 
- the nuclear stopping power -, lost for the radiative transitions. Both stopping powers can be quantitatively predicted, \emph{e.g.} by using Ziegler tables \cite{Ziegler} 
appealing to the work of Lindhard \emph{et al.} \cite{LSS}. Within the \emph{INDRA} collaboration, we use stopping power tables 
 for heavy ions in solids from Northcliffe and Schilling at low energies \cite{Northcliffe} and from Hubert and Bimbot at high 
energies \cite{Hubert}, both matched at 2.5 MeV/nucleon. They provide quite accurate results in the low and intermediate energy range, \emph{i.e.} from few hundreds of keV/nucleon 
up to $100$ MeV/nucleon, of interest here. More than a decade ago, P\^arlog \emph{et al.} \cite{Parlog1,Parlog2} put in evidence the role of the two types of energy loss to the 
quenching and also found the dependence of the fraction $\mathcal{F}(E)$ on the instantaneous velocity (or energy per nucleon $E/A$). They disentangled 
the contributions of the carriers produced in the main particle track and of the $\delta$-rays to the scintillation too. The authors quantified these 
processes in a simple Recombination and Nuclear Quenching Model
($RNQM$) connecting the exact value of the total emitted light
$\mathcal L$ to both the electronic and nuclear infinitesimal stopping powers along the incident particle track via numerical integration \cite{Parlog1}. The model 
contains Eq. (\ref{equ1}) as a particular case. Under well argued approximations, they derived a more friendly analytical formula relating 
$\mathcal L$ to the quantities $Z, A$ and $E_0$ \cite{Parlog2}:

\begin{equation}
\mathcal{L}(E_0) = a_1 E_0\left[1-a_2 \frac{AZ^2}{E_0}\ln\big(1+\frac{1}{a_2 AZ^2/E_0}\big)+a_2 a_4 \frac{AZ^2}{E_0}\ln\big(\frac{E_0+a_2 AZ^2}{a_3A+a_2 AZ^2}\big)\right],  
\label{equ3}
\end{equation}
for an incident energy $E_0$ in the CsI(\emph{Tl}) higher than the threshold $E_\delta$ at which the $\delta$-rays start to be generated. Besides the 
coefficients $a_1, a_2$, with the same physical signification as above in Eq. (\ref{equ2}), two others appear: the energy per nucleon $a_3=e_\delta$, (a few MeV/nucleon) 
and $a_4 = \mathcal {F}$ - the fraction (a few tenths of percents) of energy - they are carrying off, taken as a constant irrespective of current energy $E$ along 
the particle path above $E_\delta$. At low energy ($E \leq E_\delta$), $\mathcal{F} = 0$ and only the first term 
is present, then Eq. (\ref{equ3}) is reduced to Eq. (\ref{equ2}). These four parameters have then to be evaluated by using a number of suitable calibration 
points by a fit procedure.

The relation (\ref{equ3}) is purely analytical and can then be easily implemented for calibration purpose. It is less accurate than the 
exact treatement provided in $RNQM$ \cite{Parlog1} especially at low energy.
One drawback is also the step function used for $\mathcal {F(E)}$, 
which jumps from 0 to $a_4$ at $E = E_\delta$ in order to allow the analytical integration over $E$. This introduces a discontinuity in the function 
$\mathcal{L}(E)$ at this connection point, especially for very heavy fragments \cite{Parlog2}. Nevertheless, it may be \emph{ad hoc} improved by slightly 
improving the continuity of the fraction $\mathcal {F}(E)$ around $E_\delta$. In this work, we consider that the use of the analytical expression 
will only marginally affect the results, taking into account the
intrinsic quality of the Silicon wafers and of the CsI crystals of the \emph{INDRA} 
telescopes, which does not secure the precision required to appreciate such discrepancies. Moreover, the total light $\mathcal{L}$ emitted by the 
CsI(\emph{Tl}) scintillators is not directly measured, but reconstructed, through the procedure described by P\^arlog \emph{et al.} \cite{Parlog2}, starting from two components of 
the scintillation light measured by integrating the signal in the fast and slow time gates \cite{Parlog2}. Nevertheless, for a more rigorous and accurate treatment, the use of the exact 
formulation of $RNQM$ \cite{Parlog1} is preferable when possible, for example with high-quality detectors such as \emph{FAZIA} Si-CsI telescopes. This will be the subject of a 
forthcoming paper.

\begin{figure}[h!]
\begin{center}
\includegraphics[width=12cm]{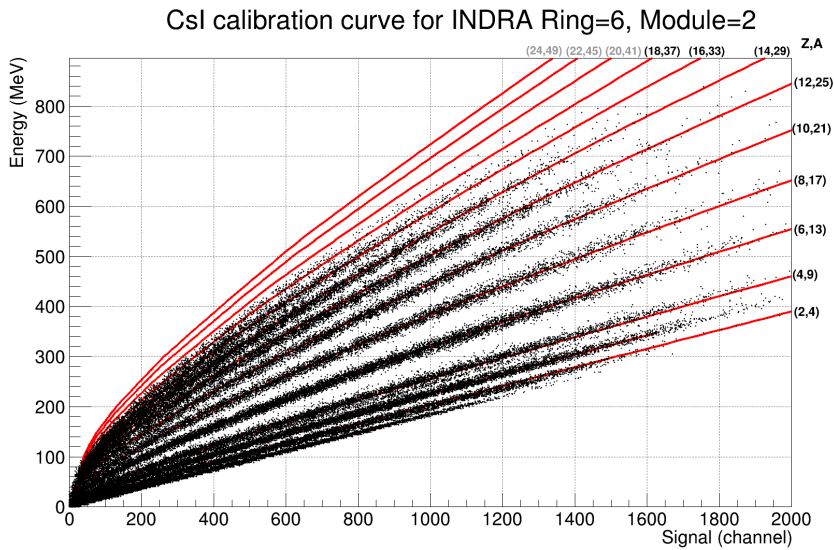}
\caption{Energy ($E_0$) - Light ($\mathcal L$) correlation in a CsI(\emph{Tl}) scintillator, for even-Z fragments. The points correspond to data from the telescope $2$ of 
the ring $6$ ($14$\textdegree $\le \theta \le 20$\textdegree) of the \emph{INDRA} 4$\pi$ array for the system $^{136}Xe+^{124}Sn$ at $32A$ MeV. The atomic 
number of the heaviest fragments emitted in this angular domain is $Z=16$. The full/coloured lines illustrate the Eq. (\ref{equ3}) predictions. The mass 
number for each element is likely corresponding to the most probable isotope, here $A=2Z+1$. See text for explanation.}
\label{fig01}
\end{center}
\end{figure}

As an example of the quality attained with our analytical description for the scintillation light in CsI(Tl) crystal, Fig. \ref{fig01} displays the energy-light correlation 
$E_0-\mathcal{L}$ using Eq. (\ref{equ3}) superimposed on \emph{INDRA} data concerning the system $^{136}Xe+^{124}Sn$ at $32A$ MeV, for a specific Si-CsI 
telescope. 

Each full/coloured line in Fig. \ref{fig01} corresponds to a given nucleus with an atomic number $Z$ and a mass number $A$.
We have chosen here to display 
isotopic lines with $A=2Z+1$ for even-Z nuclei. We will see in the following that this mass assumption is quite reasonable for IMFs when considering the 
neutron (n)-rich system $^{136}Xe+^{124}Sn$. For a given energy $E_0$ 
- determined as shown in the next section -, 
the heavier the nucleus, the smaller the light value $\mathcal L$ is; 
this is a direct consequence of the ratio nuclear/electronic stopping powers, and also of the quenching effect. Both quantities increase with the 
charge and the mass of the fragment and decrease when $E_0$ increases. Additionaly, above a certain velocity, 
$\delta$-electrons are generated, very efficient for light production. These are the reasons why the curvature of the different isotopic curves 
shown in Fig. \ref{fig01} evolves toward a linear behavior at higher light/energy, here $\mathcal{L}>600$. It is worthwhile to mention that the $\delta$-rays contribution to the light 
is quite large, reaching  $20-50\%$ for $Z>20$, as pointed out in Ref. 
\cite{Parlog2} and must be definitely included in order to reproduce the
experimental data. 
To obtain the results displayed in Fig. \ref{fig01}, we have used calibration points coming from secondary light beams stopped in CsI detectors 
from $Z=1$ up to $Z=5$ together with punched through events in the Silicon layer when possible. In a two 
dimensional $\Delta E - \mathcal L$ plot, these points are close to the ordinate $\Delta E$ axis, \emph{i.e.} to fragment energies slightly higher than that necessary to traverse 
the Silicon stage of the telescope and to reach the CsI(Tl) one with a quite small residual energy, sufficient however to be seen in the 
scintillator stage.

\begin{figure}[h!]
\begin{center}
\includegraphics[width=13cm]{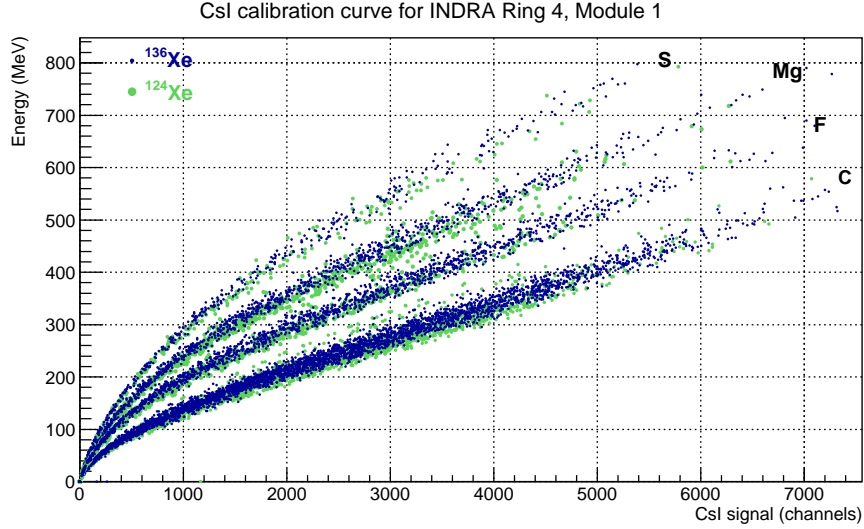}
\caption{Energy ($E_0$) - Light ($\mathcal L$) correlation in the CsI(Tl) scintillator of telescope $1$ of the ring $4$ ($7$\textdegree $\le \theta \le 10$\textdegree), 
 shown here for the two systems $^{124}Xe+^{112}Sn$ and $^{136}Xe+^{124}Sn$ at $32A$ MeV, and for some selected elements: carbon, fluorine, magnesium and sulphur. 
The large/green symbols correspond to the (n)-poor system ($^{124}Xe+^{112}Sn$) and the small/blue ones to the (n)-rich system ($^{136}Xe+^{124}Sn$).}
\label{fig02}
\end{center}
\end{figure}

In order to better appreciate the performances concerning the isotopic identification in \emph{INDRA} CsI telescopes, we display in Fig. \ref{fig02} the correlation between the 
energy and the CsI light signal (same as for Fig. \ref{fig01}) for 4 selected elements (carbon, fluorine, magnesium and sulphur), and for systems with 
different neutron content: $^{124}Xe+^{112}Sn$ and $^{136}Xe+^{124}Sn$ at $32A$ MeV. We can observe a significative difference between the two systems concerning the neutron richness 
of the produced fragments (higher masses for the (n)-rich system in blue) as one could expect from simple physical arguments. It is worthwhile to mention that this result requires 
indeed a very good stability for the CsI 
light response. This is done in \emph{INDRA} by monitoring a laser pulse all along the data taking \cite{Parlog2}. Thus, Fig. \ref{fig02} suggests that the CsI 
light signal can help to discriminate the different isotopes, here at least up to $Z=16$ (sulphur). In the following, we will use this additional valuable information to improve the 
usual $\Delta E - \mathcal{L}$ identification method for heavier elements than typically done up to carbon or nitrogen. 

\begin{figure}[h!]
\begin{center}
\includegraphics[width=13cm]{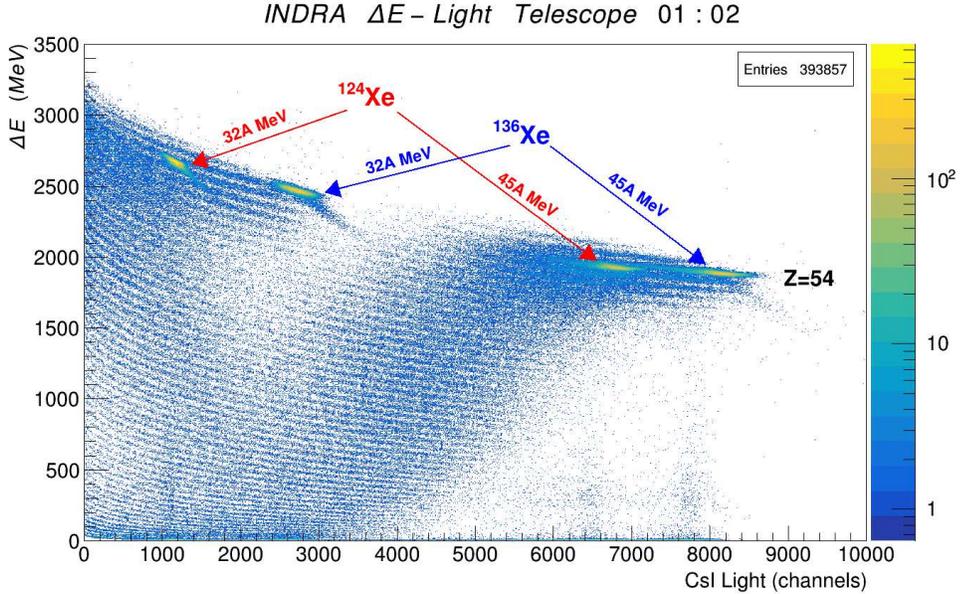}
\caption{$\Delta E - \mathcal L$ correlation for \emph{INDRA} data obtained with a module placed at a forward angle for the $^{124,136}Xe+^{124}Sn$ systems at 
$32A$ MeV and $45A$ MeV bombarding energies, illustrating its sensitivity to the detected fragment mass. The Silicon stage is $300~\mu m$ thick. 
See text for explanation.} 
\label{fig03}  
\end{center}
\end{figure}

To illustrate the overall sensitivity of the Si-CsI(Tl) telescopes to the mass number, Fig. \ref{fig03} displays the $\Delta E - \mathcal L$ correlation bidimensional matrix of the $2^{nd}$ 
module (including a $300~\mu m$-thick Si) for the $1^{st}$ ring ($2$\textdegree $\le \theta \le 3$\textdegree) of \emph{INDRA}, and for $^{124}Xe$ and $^{136}Xe$ projectiles on 
$^{124}Sn$ at $32A$ MeV and $45A$ MeV bombarding energies. The bright/yellow spots, indicated by arrows on the borders of the geometrical 
loci for $Z = 54$ (two in the region of the $32A$ MeV incident energy and other two in that of $45A$ MeV one) correspond in both cases to the (n)-poor or (n)-rich 
projectiles, respectively. These findings indeed show the good sensitivity of the response of \emph{INDRA} telescopes to the mass of the detected ejectile, thus calling for  
a deeper analysis of the experimental data as presented hereafter.

\section{Advanced Mass Estimate (\emph{AME}) in \emph{INDRA} Si-CsI telescopes} \label{INDRA}

In this section, we are going to present the new \emph{AME} identification method in details. 
We use information given by the energy lost in the Silicon detector, $\Delta E$, and the atomic number $Z$ taken from the usual $\Delta E-\mathcal L$ 
identification method in a Si-CsI map (see Fig. \ref{fig03} for example). Doing so, we benefit from the previous identification works done for \emph{INDRA} data : $Z$ 
identification in Si-CsI matrices by semi-automatic \cite{Tassan-Got} or handmade grids and the careful calibration of the Silicon detector, by means of 
$\alpha$ particle source and secondary beams stopped in this layer \cite{Tabacaru}. For heavy ions ($Z>15$), the Pulse Height Defect (PHD) in this detector can 
be large \cite{Tabacaru} and has to be carefully evaluated. For \emph{INDRA}, we use the elastic scattering of low-energy heavy ion beams (Ni and Ta at $6$ AMeV) 
which are stopped in Silicon detectors. Traditionally, we quantify the PHD as a function of the atomic number, the energy of the particle and the quality of the 
detector, according to Moulton formula \cite{Moulton}.

For a given element characterized by its atomic number $Z$, the measured energy $\Delta E$ deposited in the first layer of the Si-CsI(Tl) telescope depends on the velocity, 
or the initial energy and the mass of the incident particle 
and, in principle, it can not provide by itself the two quantities without ambiguity. To perform consistently the isotopic identification in Si-CsI matrices, we then 
assume a starting value $A_0$ for the mass number concerning one detected
nucleus with its atomic number $Z$ and, by constraining the energy loss $\Delta E$ in the Silicon stage at the 
measured value, we compute both the total energy at the entrance of
the Silicon stage and the residual energy $E_0$ deposited in the CsI(\emph{Tl}) by using the above-mentioned range and energy loss tables \cite{Northcliffe,Hubert}. This procedure imposes 
also to accurately evaluate the thickness of the $\Delta E$ Silicon detector. The value of the scintillation light $\mathcal{L}$ given by Eq. (\ref{equ3}) for the residual energy value 
$E_0$ associated to this starting value 
of $A$ is then compared to the experimental light output $\mathcal L_{exp}$ from the CsI(\emph{Tl}). In order to determine the best 'theoretical' value 
$\mathcal L(E_0)$, we iterate on mass number $A$ (and consequently on the value of $E_0$) until we find the best agreement between the theoretical and 
experimental values of the light, always compatible with the energy lost in the Silicon stage. It is worthwhile to mention that the mass number is an 
integer and, as such, is varied by increment of one mass unit. At the end of the iteration, we get an integer mass number, giving  
the best agreement for the experimentally determined quantities $\Delta E$ and $\mathcal L_{exp}$ as displayed in Fig. \ref{fig03}. This is the basis of the Advanced 
Mass Estimate (\emph{AME}) method, which, by making use in a consistent way of the experimental quantities $\Delta E$ and $\mathcal{L}_{exp}$, brings a more accurate information 
on both the mass and the residual energy (and consequently the total energy too).

As one may guess from Fig. \ref{fig03}, the calibration for the Silicon detector should be as accurate as possible to perform the best isotopic identification. 
The formula given by \cite{Moulton} used to calculate the PHD does not depend on the ion mass. This is certainly an advantage as it simplifies our approach, but it 
may become a drawback too. Even if the calibration of the Silicon stage can be considered to be rather accurate, we estimate that it represents at present 
time one of the known limitations for the extension of the identification method toward heavier nuclei ($Z>30$). Nevertheless, we will see in the following that it does not 
 hamper very much the isotopic identification for such heavy products. 

In the next sections, we will estimate the performances of this new identification procedure. As a first step, we will benchmark 
the new method for light nuclei where isotopic identification is already achieved ($1 \le Z \le 6-8$) with traditional methods. In a second step, we will then get some 
quantitative values concerning the improvement for the isotopic identification of heavier nuclei, up to xenon isotopes in our case. 

\subsection{Benchmark with the standard $\Delta E -\mathcal{L}$ method}

\begin{figure}[!h]
\begin{center}
\includegraphics[width=10cm]{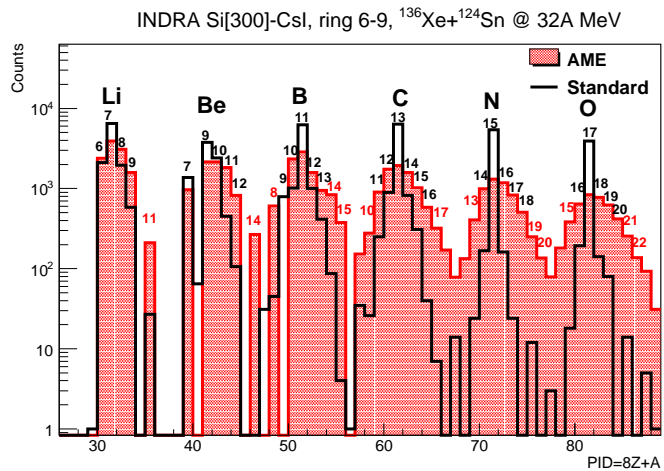}
\caption{Isotopic distributions from lithium (left) to oxygen (right). The isotopes are indicated by their mass numbers. The black and grey/red labels are explained in the text. 
Data concern the rings 6 to 9 for the 
system $^{136}Xe+^{124}Sn$ system at $32A$ MeV recorded with \emph{INDRA} and correspond to forward angles between $14$\textdegree and $45$\textdegree in laboratory 
frame. Empty histograms are the results obtained with the standard $\Delta E-\mathcal{L}$ method and filled 
 histograms those from the \emph{AME} method for the same sample of events. The Silicon stage is $300~\mu m$ thick.}
\label{fig04}
\end{center}
\end{figure}

Using the \emph{AME} method, we obtained isotopic distributions of light nuclei that have been compared to the ones obtained with the standard method (making use of  
semi-automatic \cite{Tassan-Got} or handmade grids) for \emph{INDRA} Si-CsI telescopes. Fig. \ref{fig04}
displays the isotopic distributions obtained by the new \emph{AME}
method (filled histograms) 
and the standard $\Delta E-\mathcal L$ one, using standard grids (empty histograms), from lithium ($Z=3$) up to oxygen ($Z=8$) isotopes. The numbers indicate the isotope masses. For 
the meaning of the colours of these numbers : black or gray/red, see section (IV). The Particle IDentifier (PID) defined as 
$PID=8Z+A$, and allowing to separately observe the neighbouring elements, was chosen as abscissa for this representation. The modules incorporating  
silicon detectors of only $300~\mu m$ thickness were kept for this representation. We observe an overall good agreement for the most probable isotopes, found as having the 
mass number $A = 2Z+1$ as already discussed for Fig. \ref{fig01}. 

We also notice that the new method can still
provide isotopic identification for less abundant species ((n)-rich and (n)-poor carbon to oxygen isotopes for example)
since it does not use any visual recognition to build the grids for which a sufficiently large production cross-section is needed. 
This is clearly an improvement compared to the standard methods since it allows to recover the overall isotopic distributions for a
given element $Z$, at least in this range of atomic numbers $Z=3-8$. This new feature is welcome for studying isospin effects as for example isotopic yields 
or isoscaling \cite{WCI, Ademard}. 

\begin{figure}[!h]
\begin{center}
\includegraphics[width=10cm]{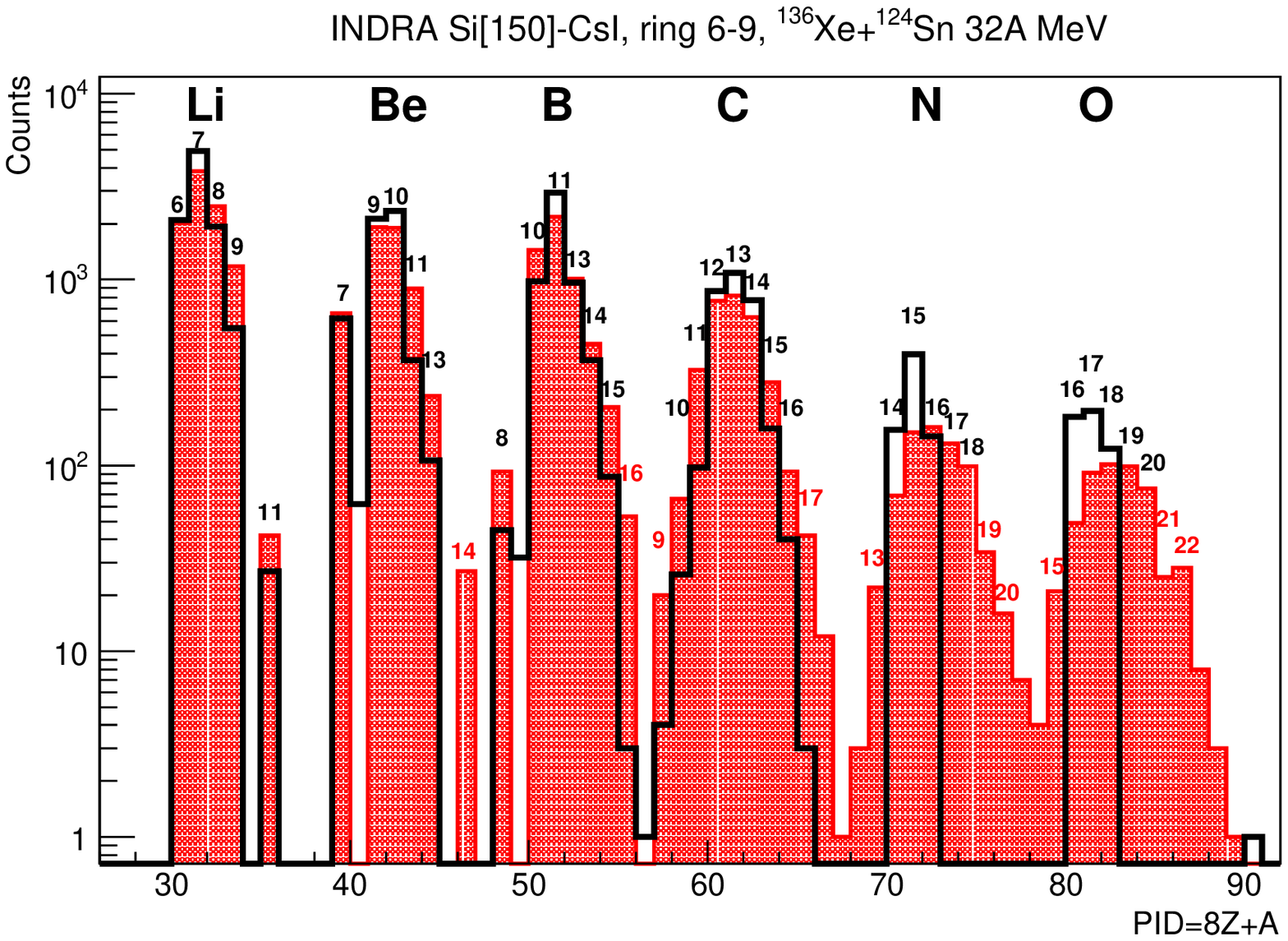}
\caption{Isotopic distributions from lithium (left) to oxygen (right). The isotopes are indicated by their mass numbers. The black and grey/red labels are explained in the text. 
Data concern the specific telescopes 
with a high gain $150~\mu m$ Silicon from rings 6 to 9 for the system $^{136}Xe+^{124}Sn$ system at $32A$ MeV recorded with \emph{INDRA} and correspond to forward 
angles between $14$\textdegree and $45$\textdegree in the laboratory frame. Empty histograms are the results obtained with the
standard $\Delta E-\mathcal{L}$ method and filled histograms those from the \emph{AME} method for the same sample of events.}
\label{fig05}
\end{center}
\end{figure}

To complete the benchmark on light nuclei, we also present in Fig. \ref{fig05} the 
isotopic distributions obtained for the specific $150~\mu m $-thick
Silicon detectors with a high gain, but for lower statistics. These ones allow to better 
discriminate the isotopes for light IMFs (up to $Z\approx8$) and constitute a more stringent test for the comparison. Actually, the mass distribution for 
the carbon isotopes given by the standard method becomes now significantly larger, closer to that provided by the \emph{AME} method, which recovers more exotic species. 
We also notice that even the yield for the most probable isotopes given by the two methods are sometimes not the same, due to the absence of grids for some telescopes where the 
visual inspection does not permit to define properly the isotopes curves and boundaries. This is particularly true for $Z=7-8$. In Figs. \ref{fig04} and \ref{fig05}, 
the black numbers indicate the masses estimated with an uncertainty lower than one mass unit, while the grey/red ones, 
those affected by higher uncertainty. This specific point is developed in section (IV).

\subsection{Comparison with different isospin systems}

To extend and confirm the previous results, we checked the isotopic identification by means of the \emph{AME} method for two systems with different isospins : $^{124}Xe+^{112}Sn$ 
and $^{136}Xe+^{124}Sn$ at the same incident energy per nucleon of $45$ AMeV without any event selection except here a common trigger multiplicity $M=1$. 
These latter are also part of the data extracted from the $5^{th}$ \emph{INDRA} campaign performed at GANIL. We could reasonably expect an overproduction 
of (n)-rich isotopes in the case of the (n)-rich $^{136}Xe+^{124}Sn$ system, for light nuclear fragments. In the 
following, we compared the isotopic distributions obtained for both systems, in order to see whether we observe any difference reflecting the possible different 
production yields for a given element $Z$.

\begin{figure}[h!]
\begin{center}
\includegraphics[width=12cm]{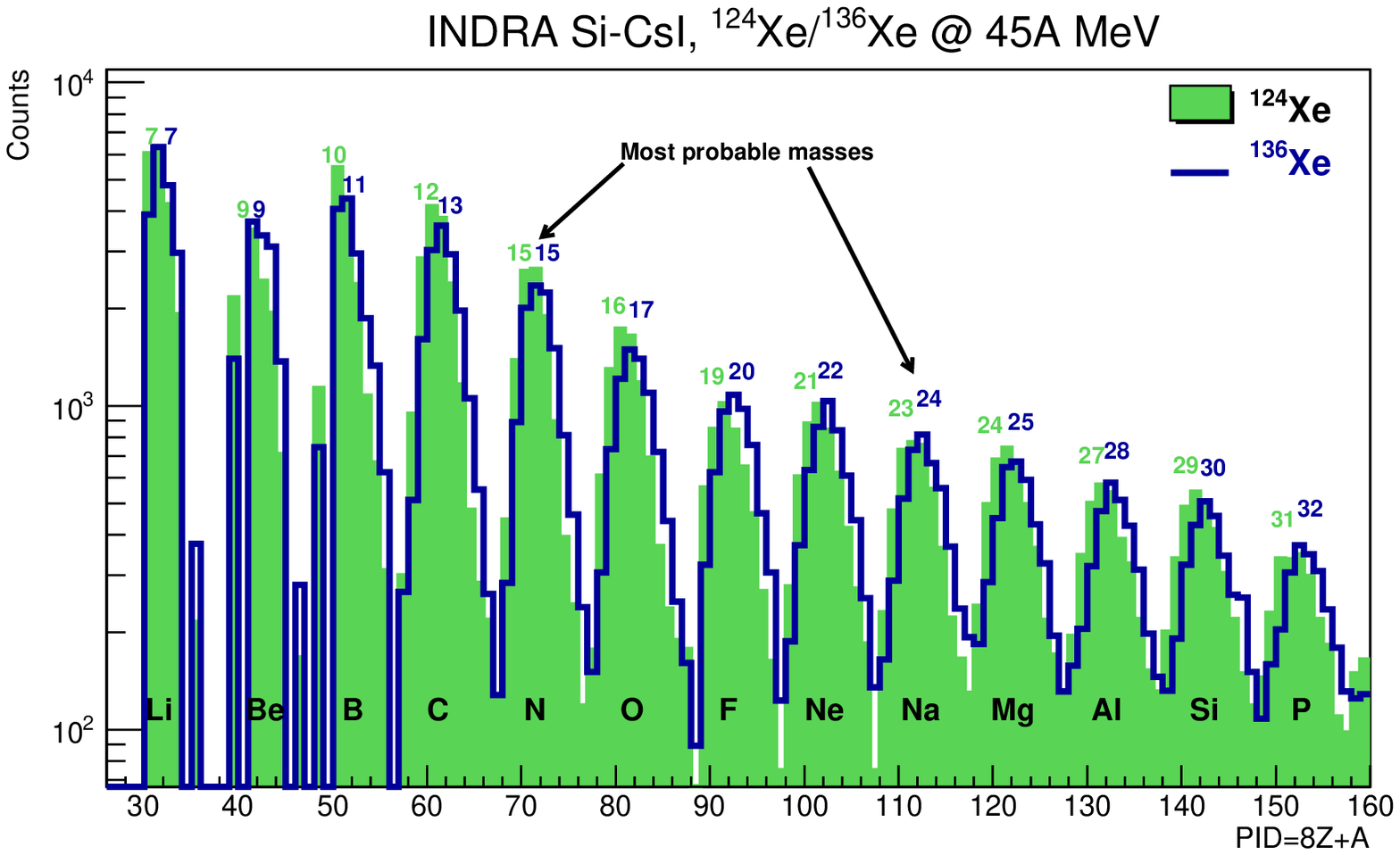}
\caption{Isotopic distributions (PID as $8Z+A$) from lithium (left) to phosphorus isotopes (right), provided by \emph{AME} with ordinary telescopes 
($300~\mu m$-thick Si). The filled histograms correspond to the $^{124}Xe+^{112}Sn$ system and the empty histograms to the $^{136}Xe+^{124}Sn$ system at 
the same incident energy per nucleon $E/A=45~MeV$ (\emph{INDRA} data). The most abundant isotope in each case is indicated too.}
\label{fig06}
\end{center}
\end{figure}

Fig. \ref{fig06} shows the isotopic distributions from lithium to phosphorus isotopes, provided by Silicon-CsI telescopes. We display here only the results for the $300~\mu m$-thick 
Silicon ones. We can observe a global shift of the isotopic distribution toward more (n)-rich species for the (n)-rich system ($^{136}Xe+^{124}Sn$) as compared to the (n)-poor one 
($^{124}Xe+^{112}Sn$). If we consider the most abundant isotope per element, it is $^{7}Li$ instead of $^{6}Li$ and $^{17}O$ instead of $^{16}O$ for example, together with the enhancement 
of very (n)-rich isotopes production for the (n)-rich system ($^{136}Xe+^{124}Sn$) as one could expect. This illustrates the fact that the isotopic distributions determined with 
$\emph{AME}$ are not an artifact of the method but they truly could be associated to the genuine (physical) isotopic distributions. 

\section{Qualifying the isotopic identification}

The isotopic identification can be further qualified by some specific operations. More precisely, we can provide a quite accurate estimate for the mass number 
$A$ even if the full isotopic resolution is not achieved. We remind that, knowing the thickness of the Silicon detector ($300~\mu m$ all along 
this section), the atomic number $Z$ value of a detected fragment and the well determined $\Delta E$, corrected for the PHD \cite{Tabacaru,Moulton}, we can 
start by proposing an atomic mass $A_i$ number and compute the corresponding residual energy $E_{0i}$ in the CsI stage using the energy loss tables. They are 
connected to the calculated scintillation light $\mathcal L_i = \mathcal L(E_{0i},A_i)$ via the Eq.(\ref{equ3}). Then, the integer mass number is varied, by steps 
of one unit, in order to minimize the quantity $|\mathcal L_i-\mathcal L_{exp}|/\mathcal L_{exp}$, in such a way that the measured $\Delta E$ value be reproduced 
too. After a few iterations, the best integer value $A^*$ of $A_i$ and the related value of $E_{0i}$ are found, characterized by the shortest normalized distance 
$d_i=|\mathcal L_i-\mathcal L_{exp}|/\mathcal L_{exp}$ between the calculated $\mathcal L_i$ and experimental $\mathcal L_{exp}$ light. Finally, to get a representative value 
$A_{est}$ for the estimated mass number in a single event (one experimental point in the $\Delta E - \mathcal L$ plot) at a given $\Delta E$, we simply weight 
the different $A_i$ values by the inverse of $d_i^2$ as:

\begin{equation}
A_{est} =  \frac{1}{\Sigma_i \frac{1}{d_i^2}}\Sigma_i \frac{A_i}{d_i^2}
\label{eq3}
\end{equation} 

We shall exploit thus not only the mass number as an integer but directly the $PID$, defined above as: $PID=8Z+A_{est}$, by letting now the mass number $A_{est}$ 
to be a \emph{real} number. Of course, for an experimental light $\mathcal L_{exp}$, the main contributions to $A_{est}$ are coming from the shortest distances $d_i$. 
Doing so, we can obtain an estimation concerning the uncertainty  $\Delta A$ by taking the absolute difference between the optimum value $A^*$, corresponding 
to the smallest distance $d_i$ and the weighted value $A_{est}$ obtained with the Eq. (\ref{eq3}): 

\begin{equation}
 \Delta A=|A^*-A_{est}|
 \label{eq4}
\end{equation}

If the two values $A_{est}$ and $A^*$ are close enough ($\Delta A<0.5$, so comprised in one unit range), we assume a full isotopic identification, whereas if $\Delta A \ge 0.5$, 
we have only a limited isotopic identification. This procedure can be therefore considered as a simple and easy way to qualify the isotopic identification. 
This is illustrated by the black and grey/red numbers on Figs. \ref{fig04} and \ref{fig05}. The black numbers refer to $\Delta < 0.5$ whereas the grey/red ones to $\Delta A \ge 0.5$.

\begin{figure}[h!]
\begin{center}
\includegraphics[width=7cm]{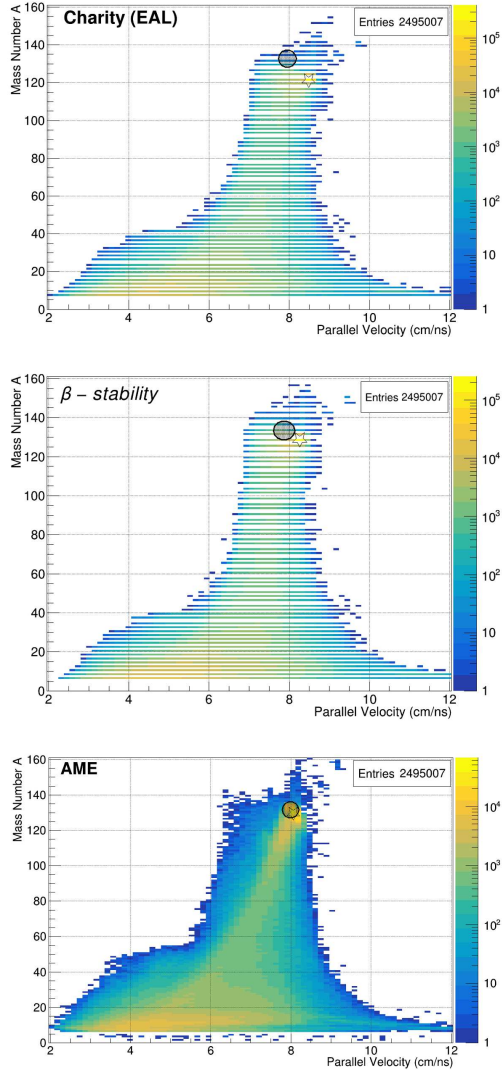}
\caption{\emph{INDRA} data. Correlation between the velocity $V_z$ in cm/ns, parallel to the beam direction, and the atomic mass $A$ for $^{136}Xe+^{112}Sn$ at $32A$ MeV 
obtained in the very forward rings (1-5) for a trigger multiplicity $M \ge 1$. The upper panel displays the traditional $\Delta E - \mathcal L$ identification method with 
no mass determination and the EAL hypothesis for the masses of the fragments. The same for the middle panel, but with the $\beta$-stability valley hypothesis for the 
masses of the fragments. In the lower panel are plotted the results obtained with the new \emph{AME} method. The circles indicate the projectile mass ($A=136$) and the parallel 
projectile velocity ($V_z=7.9~cm/ns$), while the stars correspond to the peak of each two-dimensional distribution.}
\label{fig07}
\end{center}
\end{figure}

In order to further evaluate the validity of the method, we have also used \emph{INDRA} results for the four different systems: $^{124,136}Xe+^{112,124}Sn$ 
at $32$ AMeV. Several tests are then proposed in the following. First, we have looked to data at the most forward angles, from rings $1 - 5$, \emph{i.e.} $2$\textdegree 
$\le \theta \le 15$\textdegree. These ones are obtained from the system $^{136}Xe+^{112}Sn$ at $32$ AMeV, by requiring a trigger multiplicity (fired telescopes) $M \geq 1$, in order 
to select mostly quasi-elastic events. From Fig. \ref{fig03}, we could indeed notice that we recover as main contribution the quasi-projectile ($Z \approx 54$) in the most forward rings. 
For such high $Z$ values, the isotopes are not visually separated in the $\Delta E - \mathcal L$ matrix; in the framework of the standard method, a hypothesis on the mass has to be 
made for finding their velocities starting from the measured energies deposited in the $1^{st}$ stage of a telescope. Fig. \ref{fig07} displays the correlation between the 
mass and the velocity parallel to the beam, for different ejectiles and for different mass estimates. The yellow stars indicate the maximum number of entries.

The two upper panels of Fig. \ref{fig07} display the $A-V_{z}$ correlation for the standard case (usual $\Delta E - E$ method), where only the atomic number $Z$ is determined 
from the $\Delta E - \mathcal L$ plot. In the upper panel, for $Z=54$ we took as mass hypothesis the prediction $A=120$ for the evaporation attractor line (EAL) \cite{Charity} (see below). 
It leads to value $V_{z}\approx8.5~cm/ns$ of velocity parallel to the beam direction. In the middle panel, the $\beta$-stability hypothesis is used , and the mass for xenon is set to 
$A=129$, richer in neutron, and $V_{z}\approx8.2~cm/ns$. In both cases, the values of the atomic mass $A$ and the parallel velocity are peaked quite far from the expected 
elastic contribution, in the present case: $A=136$ and $V_{z}=7.9~cm/ns$ represented by the filled circles on Fig. \ref{fig07}. This is due to the incorrect values of the mass number $A$, 
simply calculated from the atomic numbers $Z$ via different hypotheses. Consequently, the corresponding parallel velocities $V_z$ are also incorrect since they were computed by means 
of these hypothetical $A$ values. At variance, we can notice that applying the iterative $AME$ method - lower panel of Fig. \ref{fig07} -, the plotted distribution presents at 
$A\approx133$ and $V_{z}\approx8.0~cm/ns$ a maximum located much closer to the elastic contribution. We could therefore infer that the obtained results with $AME$ are more valid   
for the (n)-rich projectiles even for these very heavy ions detected in the region of quasi-elastic events. We also found the same conclusion for the proton (p)-rich system 
$^{124}Xe+^{112}Sn$ at $32A$ MeV.

\begin{figure}[h!]
\begin{center}
\includegraphics[width=13cm]{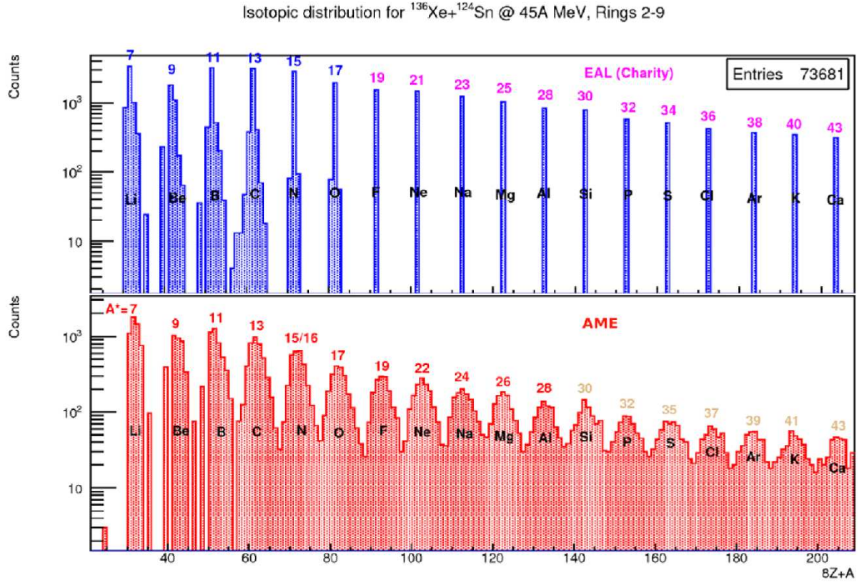}
\caption{\emph{INDRA} data. Particle IDentification $PID=8Z+A$ for the  $^{136}Xe+^{112}Sn$ at $32A$ MeV obtained in the forward rings (1-5) for a trigger multiplicity 
$M \ge 1$. The upper panel refers to the standard method: above $Z = 8$, there is no isotopic identification and a mass hypothesis (here EAL \cite{Charity}) is necessary. 
The lower panel illustrates the new \emph{AME} method, leading - for each element - to their distribution of isotopes, the most abundant ones 
being indicated by their mass numbers. The grey numbers correspond to elements where $\Delta A > 0.5$ (see text for details).}
\label{fig08}
\end{center}
\end{figure}

Now, if we look at the $PID$ distributions in Fig. \ref{fig08}, we may stress also the differences. In the upper panel, when we are not using the scintillation light 
to determine the mass number (usual method), we can get some isotopic identification up to $Z=6-8$. By contrast, as shown in the lower panel, thanks to the new \emph{AME} method, we are 
 now able to distinguish a fair isotopic identification up to at least $Z \approx 12-13$ for which we have $\Delta A \le 0.5$ as obtained from Eq. (\ref{eq4}) for the most 
abundant isotopes. The corresponding improvement concerning the isotopic resolution is indeed obtained by taking into account the additional information from the CsI crystal. 

\begin{figure}[h!]
\begin{center}
\includegraphics[width=10cm]{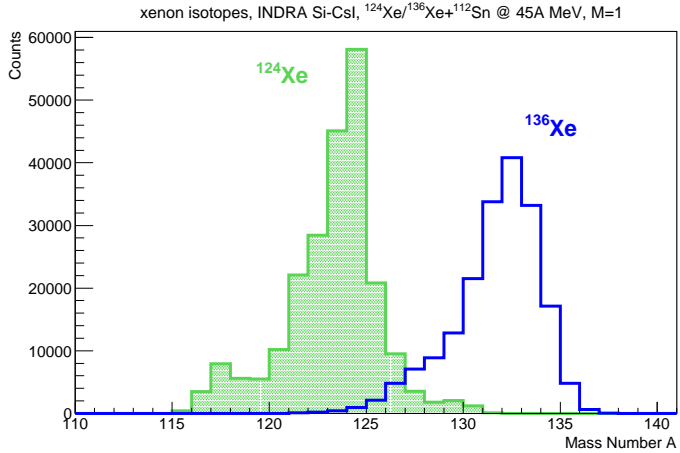}
\caption{\emph{INDRA data}. Isotopic distribution of xenon isotopes ($Z=54$) for the $^{124}Xe+^{112}Sn$ (filled histogram) and $^{136}Xe+^{112}Sn$ (empty histogram) at $45A$ MeV 
provided by the \emph{AME} method, in the most forward rings (1-3) for a trigger multiplicity $M=1$. }
\label{fig09}
\end{center}
\end{figure}

We could also qualify the accuracy of the isotopic identification for $Z$ much higher than $12$ by taking advantage once again of the elastic channel for both reactions. 
For a trigger multiplicity $M=1$, an angular range between $2$\textdegree~and $7$\textdegree~(rings 1-3), and by selecting only the xenon nuclei ($Z=54$), 
we obtain the isotopic distributions displayed in Fig. \ref{fig09}. These latters are centered around $A \approx 124$ for $^{124}Xe$ 
(the mass of the projectile) and $A \approx 133$ for $^{136}Xe$ (three mass units smaller than the projectile). For the $^{136}Xe$ data, due to its neutron richness, one 
could expect a loss of few neutrons for the projectile even in very peripheral collisions, transforming thus the elastic contribution into a quasi-elastic one. The results 
are therefore compatible with physical arguments and with those shown in the lower panel of Fig. \ref{fig07} (mass - velocity correlation). The width of these isotopic distributions 
reflect indeed the convolution of the physical isotopic distribution as well as the uncertainty on the determined mass. We can therefore reasonably deduce that 
the uncertainty $\Delta A=3$ found in Fig. \ref{fig09} could represent an \emph{upper} limit for the mass uncertainty. 

\begin{figure}[h!]
\begin{center}
\includegraphics[width=13cm]{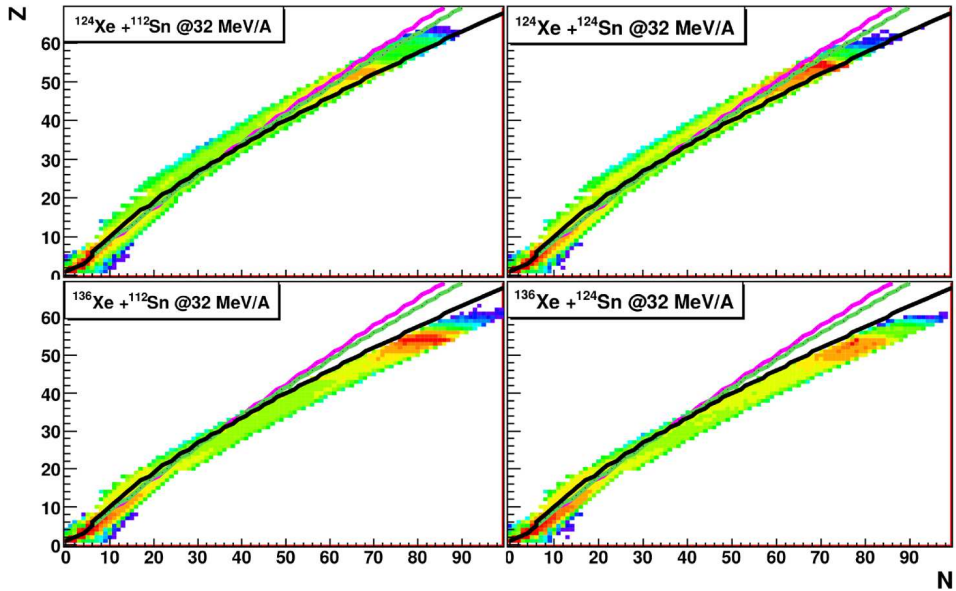}
\caption{$Z-N$ charts of the detected nuclei in the forward rings (1-9) of \emph{INDRA} for a trigger multiplicity $M \ge 4$ from the four reactions $^{136,124}Xe+^{112,124}Sn$ at 
$32A$ MeV. Data correspond to the \emph{AME} method. The different grey/coloured curves correspond to the EAL hypotheses on the neutron number $N$ and the black line to the 
$\beta$-stability valley one. See text for explanation.}
\label{fig10}
\end{center}
\end{figure}

Finally, we display in Fig. \ref{fig10} the $N-Z$ charts for the reaction products in the above-mentioned reactions, their masses being determined via the \emph{AME} 
procedure. We have selected the trigger condition $M \ge 4$, thus removing the major part of the quasi-elastic contribution presented in the previous Figs. \ref{fig07}-\ref{fig08}. 
The grey/coloured curves represent the evaporation attractor lines \cite{Charity}, predicting the number of neutrons $N$ as an integer function 
of $Z$: the steeper/pink line, with $N= 1.072Z+2.032 \times 10^{-3}Z^2$ recommended for $Z < 50$ and the more gentle slope/green line, with 
$N= 1.045Z+3.57 \times 10^{-3}Z^2$ recommended for $Z \ge 50$. The black curve indicates a $3^{rd}$ degree polynomial fit of the $\beta$-stability valley as the integer of 
$N= 1.2875+0.7622Z+1.3879 \times 10^{-2}Z^2-5.4875 \times 10^{-5}Z^3$, with \emph{i.e.} nuclei more (n)-rich than for the EAL lines. 

The $Z-N$ charts in Fig. \ref{fig10} concern the forward detection angles: $2$\textdegree $\le \theta \le 45$\textdegree (rings $1-9$ of \emph{INDRA}). These data seem 
to reflect mainly the ratio $N/Z$ of the projectile and none of these hypotheses on the number $N$ of neutrons, and consequently on $A$, is able to reproduce in average the results, 
especially for the (n)-rich $^{136}Xe$ projectiles (lower panels). This overall view pleads in favour of the \emph{AME} procedure compared to a simple mass hypothesis. 
With the present method we can obtain a better calibration of very thick CsI(\emph{Tl}) scintillators allowing at the same time the full detection of very energetic charged
 reaction products and their mass determination with the best resolution. \emph{AME} upgrades thus the $4 \pi$ \emph{INDRA} array, designed to measure only the atomic number $Z$ of 
 the heavy nuclear fragments stemming from multifragmentation reactions, to a device able to estimate their mass $A$ too, up to $Z\approx12-13$ for an isotopic resolution 
$\Delta A \le 0.5$ and $Z\approx54$ for $\Delta A \le 3$, and this in a very compact geometry.

\section{Conclusion}
 
We have presented a method called Advanced Mass Estimate (\emph{AME}), a new approach for isotopic identification in Si-CsI telescopes using the analytical formulation 
for the CsI(\emph{Tl}) light response provided in \cite{Parlog2}. It includes explicitely the light quenching and the $\delta$-rays contribution to the scintillation of the CsI(\emph{Tl}) 
crystals. In this framework, we have shown that it is possible to use an iterative procedure to accurately calibrate the CsI detectors and, at the same time, to 
estimate the mass number $A$ of the charged reaction products, besides the charge $Z$ one, with a resolution better than the one previously achieved by standard techniques.
This method allows to recover not only the isotopic distributions obtained by the usual
visual techniques for $Z=1-8$, but it can also be extended to heavier nuclei up to $Z\approx12-13$, 
with an uncertainty of one atomic mass unit for the telescopes of the \emph{INDRA} array. In addition, from the comparison with experimental data, we have shown 
that it is reasonably possible to estimate the atomic mass within $2-3$ mass units up to xenon isotopes, if one is able to carefully evaluate the thickness and the pulse height defect 
in the $\Delta E$ silicon layer. We then consider that the quality 
of \emph{INDRA} Si-CsI experimental results can be dramatically improved by using the new \emph{AME} method, and that is particularly well adapted to undergo analyses 
with radioactive beams exploring a large $N/Z$ domain. The \emph{AME} method is not only suited for \emph{INDRA} Si-CsI(Tl) telescopes but can be also 
successfully exploited with any charged particle array using the same kind of telescopes. Further studies concerning the implementation of the Recombination and Nuclear Quenching 
Model with the exact treatement mentioned in the first section are currently in progress, by using high-quality data from \emph{FAZIA} telescopes, and will be the subject of 
a forthcoming paper.

\end{document}